\documentclass{article}
\usepackage{fortschritte2}
\usepackage{epsfig}
\usepackage{amssymb}
\def\beq{\begin{equation}}                     %
\def\eeq{\end{equation}}                       %
\def\bea{\begin{eqnarray}}                     
\def\eea{\end{eqnarray}}                       
\def\ba{\begin{array}}
\def\ea{\end{array}}
\def\IZ{\mathbb{Z}}

\def\IP{\mathbb{P}}
\def\CO{{\cal O}}

\def\beq{\begin{equation}}
\def\eeq{\end{equation}}

\def\Tr{\mathop{\rm Tr}}

\def\eref#1{(\ref{#1})}

\def\nn{\nonumber}
\begin {document}                 
{\flushright{\small MIT-CTP-3456\\UPR-1060-T\\hep-th/0312222\\}}
\def\titleline{
%
%
%
A Trio of Dualities:
Walls, Trees and Cascades
}
\def\email_speaker{
{\tt 
%
%
sfranco@mit.edu, hanany@mit.edu, yanghe@physics.upenn.edu             
}}
\def\authors{
%
%
%
%
%
Sebasti\'{a}n Franco,\1ad Amihay Hanany,\1ad$^,$\2ad\sp        %
Yang-Hui He\,\3ad
}
\def\addresses{
%
%
%
%
\1ad                                        %
Center for Theoretical Physics,\\           %
Massachusetts Institute of Technology,\\ 
Cambridge, MA 02139, USA\\   
\2ad
Institute for Advanced Study, \\
Princeton, NJ 08540, USA \\
\3ad                                        
Department of Physics and Math/Physics RG,\\
The University of Pennsylvania,\\
Philadelphia, PA 19104-6396, USA		
}
\def\abstracttext{
%
%

We study the RG flow of ${\cal N}=1$ world-volume gauge theories of D3-brane 
probes on certain singular Calabi-Yau threefolds. Taking the gauge theories 
out of conformality by introducing fractional branes, we compute
the NSVZ beta-function and follow the subsequent RG flow in the 
cascading manner of Klebanov-Strassler. We study the duality trees that blossom 
from various Seiberg dualities and encode possible cascades.
We observe the appearance of duality walls, a finite
limit energy scale in the UV beyond which the dualization cascade cannot proceed.
Diophantine equations of the Markov type characterize the dual phases of these 
theories. We discuss how the classification of Markov equations for different 
geometries into families relates the RG flows of the corresponding gauge theories.


}
\large
\makefront

{\it
\noindent
Submitted to "Fortschritte der Physik" as part of the proceedings of 
the "36th International Symposium Ahrenshoop on the Theory of 
Elementary Particles," Berlin, August 26-30, 2003, and based on talks given by A. Hanany in Berlin and The University of Pennsylvannia, Philadelphia.
}

\section{Introduction and Conclusions}

D3-brane probes on singular Calabi-Yau threefolds have been widely used
both as a means to study (potentially realistic) four dimensional, ${\cal N}=1$
supersymmetric gauge theories and as a tool for the investigations of
many beautiful mathematical phenomena (for a review,
cf.~e.g.~\cite{rev}). A central phenomenon in these theories is the the 
existence of Seiberg duality \cite{Seiberg:1994pq}. This corresponds to an exact equivalence
between different gauge theories in the IR limit. Armed with various technologies, 
such as the Inverse Algorithm (which computes the gauge theories on the world-volume
of D3-branes probing toric singularities), $(p,q)$-webs, exceptional collections and the
derived category, a host of gauge theories have been constructed for a plethora of geometries
that constitute non-trivial horizons for the bulk AdS theory. Of
particular interest are the various guises and generalization of
Seiberg duality, called ``Toric Duality'' in \cite{inverse,toricdual}
(for a short introduction, cf.~e.g.~\cite{conf1}), that emerge naturally
when realizing the gauge theories geometrically in String Theory. 

The world-volume theories on the D-brane probes are quiver theories. By successive 
applications of dualities, it is possible to construct, for each probed geometry,
an infinite web of dual phases. The constructions that represent the space of dual 
theories and the connections between them have been affectionately called 
{\bf Duality Trees} and {\bf Flowers} \cite{Franco:2003ja}. 
As we dualize upon a node in the quiver at each stage, a new branch blossoms. 

A famous example which exploits Seiberg duality
and which exhibits many intricacies of the AdS/CFT
correspondence is the theory for the conifold \cite{klst}. With the
addition of $M$ fractional branes, the theory is taken out of conformality
and there is a subsequent RG flow for the gauge couplings. This
running can be identified with the radial 
variation of the fiveform flux in AdS. 
The gauge theory interpretation of this flow states that when one of the couplings 
becomes strong, one, \`{a} la Seiberg \cite{Seiberg:1994pq}, dualizes the theory and flow to
the IR. This RG flow can be followed {\it ad infinitum} and the process was referred 
to as a {\bf cascade}. Thus, every time one of the couplings becomes infinite, we switch to 
an alternative dual description in which the effective number of D3-branes is decreased
by $M$.

When dealing with non-conformal theories, duality trees represent the possible paths or cascades
that can be followed along an RG flow. In this context, the topology of the tree can be used
to identify the qualitative features of the possible flows. For example, the existence of 
closed cycles would signify that certain dualities may be trapped within a group of
theories, given rise to a cascade resembling the conifold one.

The generalization of the cascade phenomenon to other geometries is hindered
by the fact that the conifold is really the only geometry for which we
know the metric. Nevertheless nice extensions from the field theory
side have been performed \cite{fiol,strassler,HW}. There has been some
evidence that for other geometries the cascade is
qualitatively different. An interesting possibility is the existence of 
{\bf Duality Walls}. This concept was introduced in \cite{strassler}, where it 
appeared as fundamental limitation on the UV scale of the theory when trying 
to accommodate the Standard Model at the IR limit of a duality cascade.


Inspired by the myriad of dual phases of theories constructed
for the immediate and most important generalizations of the conifold,
namely the del Pezzo surfaces \cite{inverse}, we study their
cascades. 
We focus on the illustrative example of $F_0$, the zeroth
Hirzebruch surface. We find that there is indeed a duality wall past
which Seiberg duality can no longer be performed. To this claim
we shall soon supplant with analytic proof \cite{FHHW} and also
demonstrate some fascinating chaotic behavior.
In due course of our investigations, we shall also delve into such
elegant mathematics as resolution of singularities and the emergence
of certain Diophantine equations of Markov type
which characterize the duality tree. We will discuss how the connection 
between Markov equations for different singularities can be exploited
to relate RG cascades for these geometries.
We will see intricate inter-relations between D-brane gauge theories,
algebraic singularities, AdS holographic duals, finite graphs and
Diophantine equations.

\section{The Beta Function}

Let us study the quiver theory arriving from a stack of $N$ parallel
coincident D3-branes on a singular Calabi-Yau threefold. This is an
${\cal N}=1$ conformal 4D theory with gauge group $\prod\limits_i U(N_i)$ 
and chiral bifundamental multiplets with multiplicities $A_{ij}$
between the $i$-th and $j$-th gauge factors (the matrix $A_{ij}$ is
the adjacency matrix of the quiver). Anomaly freedom requires that the
ranks $N_i$ obey

\beq
\label{anofree}
(A - A^T)_{ij} \cdot \vec{N} = 0 \ .
\eeq

We can add $M$ fractional branes, whereby modifying the ranks $N_i$,
in such a way that \eref{anofree} is still satisfied. These new ranks
$n_i=N_i + M_i$ take the theory out of conformality, inducing a nontrivial RG flow.
This flow is governed by the NSVZ beta-function \cite{NSVZ} that for each gauge group
takes the following form
\beq
\beta_i = \frac{d(8\pi^2 /g_i^2)}{d \ln \mu}=¨
{3T(G)-\sum_i T(r_i)(1-\gamma_i) \over 1-{g_i^2 \over 8 \pi^2} T(G) }:= \frac{d x_i}{d t}  
\label{betadef}
\eeq
where $\mu$ is the energy scale; we have defined $x_i$ to be the inverse square 
gauge coupling and $t$ to be the log of the energy. For an $SU(N_c)$ gauge group,
$T(G)=N_c$ and $T(fund)=1/2$.

An interesting limit in which computations can be carried on easily, is the one in which 
the number of fractional branes is much smaller than the number of probe branes. In this regime, 
the beta functions can be computed as follows \cite{intri,chris,Franco:2003ja}. Let $\gamma_{ij}$ 
be the anomalous dimensions of the field $A_{ij}$. The beta functions for the gauge and 
superpotential couplings, can be written as

\bea
\beta_{i \in \mbox{nodes}}
&=& 3 n_i -\frac12 \sum_{j=1}^k (A_{ij} + A_{ji}) n_j + \frac12
	\sum_{j=1}^k (A_{ij} \gamma_{ij} + A_{ji} \gamma_{ji}) n_j \nn \\
\beta_{h \in \mbox{loops}} 
&=& \mbox{length}(loop) - 3 + \frac12 \sum_{h} \gamma_{h_i h_j} \ ;
\label{betasquiver}
\eea
where we consider one $h$ coupling for each of the gauge invariant terms in the
superpotential, which are represented by closed loops in the quiver.\footnote{The
number of independent superpotential couplings can be reduced in the presence of
symmetries.}
We then solve for $\gamma_{ij}$ by setting \eref{betasquiver} to zero
at the conformal point where $n_i = N_i$, and fix the remaining freedom using 
maximization of the central charge \cite{intri}
\beq
a=\frac{3}{32}(3\Tr R^3-\Tr R) =
\frac{3}{32} \left[2 \sum_i N_i^2 + \sum_{i<j} A_{ij} N_i N_j \left[
3 (R_{ij}-1)^3 - (R_{ij}-1) \right ]\right]
\eeq
where the anomalous dimensions are related to the corresponding $R$ charges 
by $\gamma_{ij}=3 R_{ij}-2$. With the solutions, back-substitute 
into the beta-functions and proceed out of conformality by taking the ranks
$n_i = N_i + M_i$. We remark that the beta function thus computed is
of leading order and we assume that the next order is $\CO(M/N)^2$
while $\CO(M/N)$ vanishes.

\section{The Conifold Cascade}

It is instructive to review our methodology by applying it to the classic 
Klebanov-Strassler conifold theory. In this case, we have
an $SU(N) \times SU(N)$ gauge theory with four bifundamental 
chiral multiplets $A_{1,2}$ and $B_{1,2}$ and superpotential as 
given in \eref{coni}

\beq
\label{coni}
\ba{c}
\epsfxsize = 4cm
\epsfbox{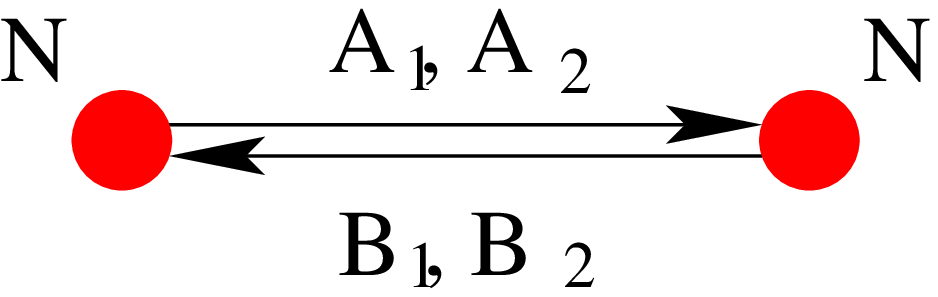}
\ea~~~~~~~~~
W = \epsilon^{ij} \epsilon^{kl} \Tr A_i B_k A_j B_l \ .
\eeq

Adding fractional branes in accordance with \eref{anofree}, 
we have an $SU(N+M) \times SU(N)$ theory and the beta functions, using
the techniques outlined above, are given in \eref{conisum}.
Indeed by the $\IZ_2$ symmetry of the quiver, the $\CO(M/N)$
corrections to the beta functions are zero. The subsequent duality
tree consists of a single node with a single closed cycle taking
the theory (up to permutation of gauge groups) back to itself. The 
running of the two beta-functions cascade alternatingly in weave pattern:

\beq\label{conisum}
\ba{cc}
\ba{rcl} 
SU(N+M): \beta_{g_1} &=& 3M \nn \\
SU(N):   \beta_{g_2} &=& -3M \ ; 
\ea
&
\ba{ccc}
\mbox{Duality Tree} & &\mbox{Running of Couplings} \\
\ba{l} \epsfxsize = .6cm \epsfbox{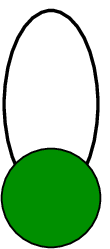} \ea
& \quad &
\ba{r} \epsfxsize = 6cm \epsfbox{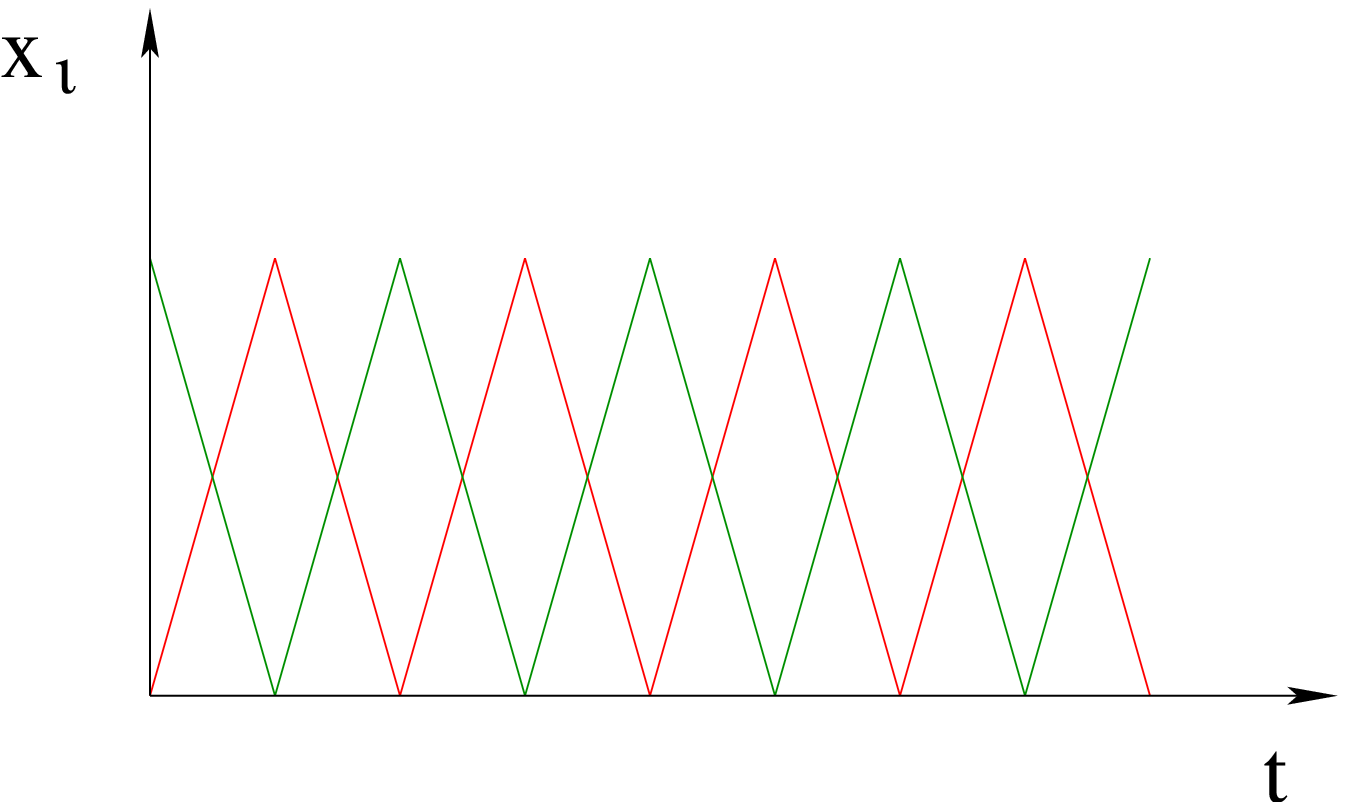} \ea
\ea\ea
\eeq

We see that the $t$ interval between consecutive dualizations is constant along the 
cascade. The ranks of the gauge group vary linearly with the step, decreasing towards 
the IR, and thus diverging as the logarithm of the energy in the UV.

\section{Cascade for $F_0$, the Zeroth Hirzebruch}

\label{section_F0_cascade}

We can now move on to the theory of our principal concern, the gauge theory 
corresponding to the probe on a complex cone over $F_0$, the zeroth Hirzebruch 
surface, or $\IP^1 \times \IP^1$. Interesting phenomena arise in this case.
The duality tree is much richer; it resembles a flower which we call
{\it flos Hirzebruchiensis}. This is shown, together with some dual
phases of the quivers in plot \eref{F0}
\beq\ba{ccc}
\ba{l} \epsfxsize = 6cm \epsfbox{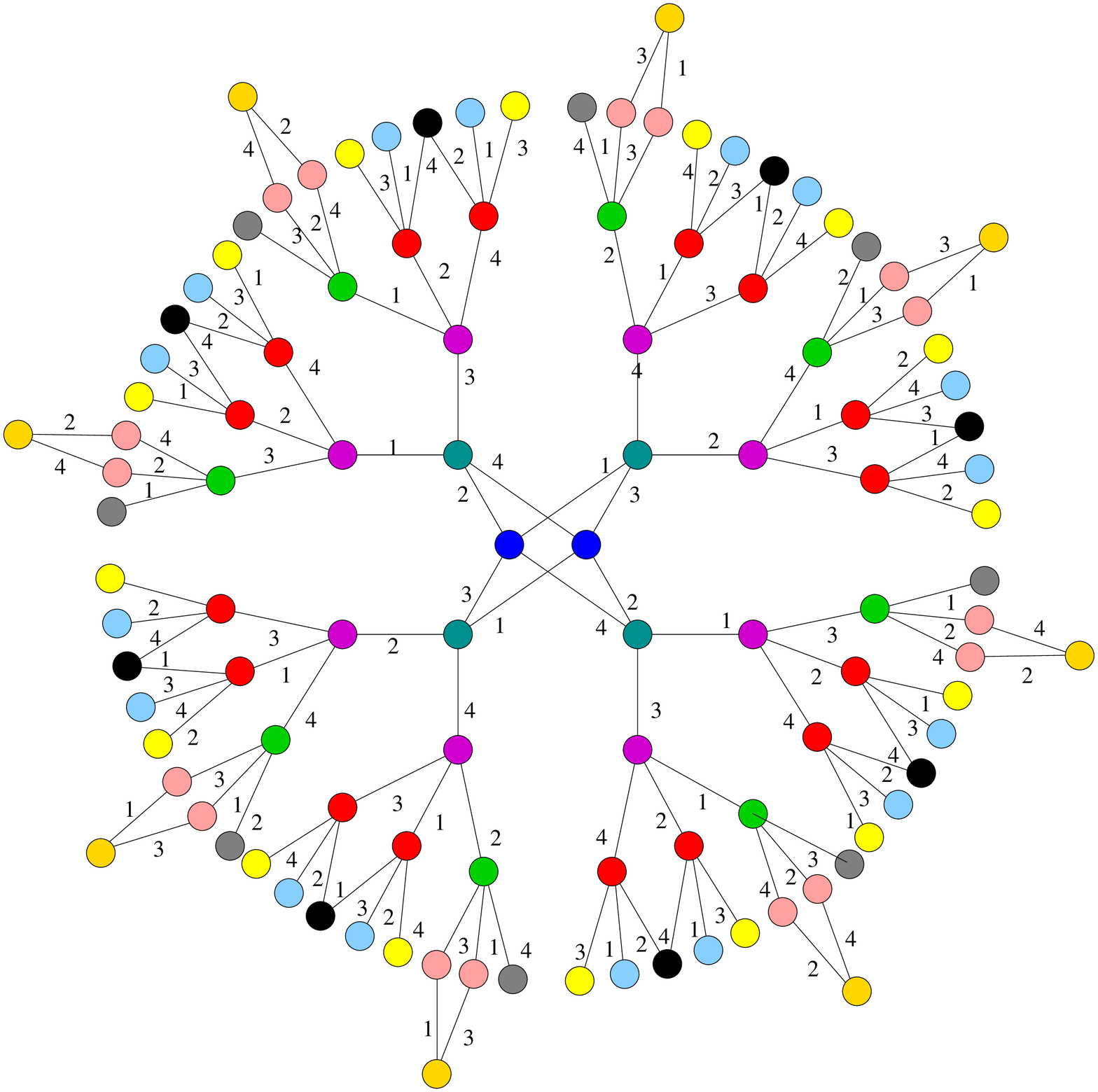} \ea
& \quad &
\ba{r} \epsfxsize = 6cm \epsfbox{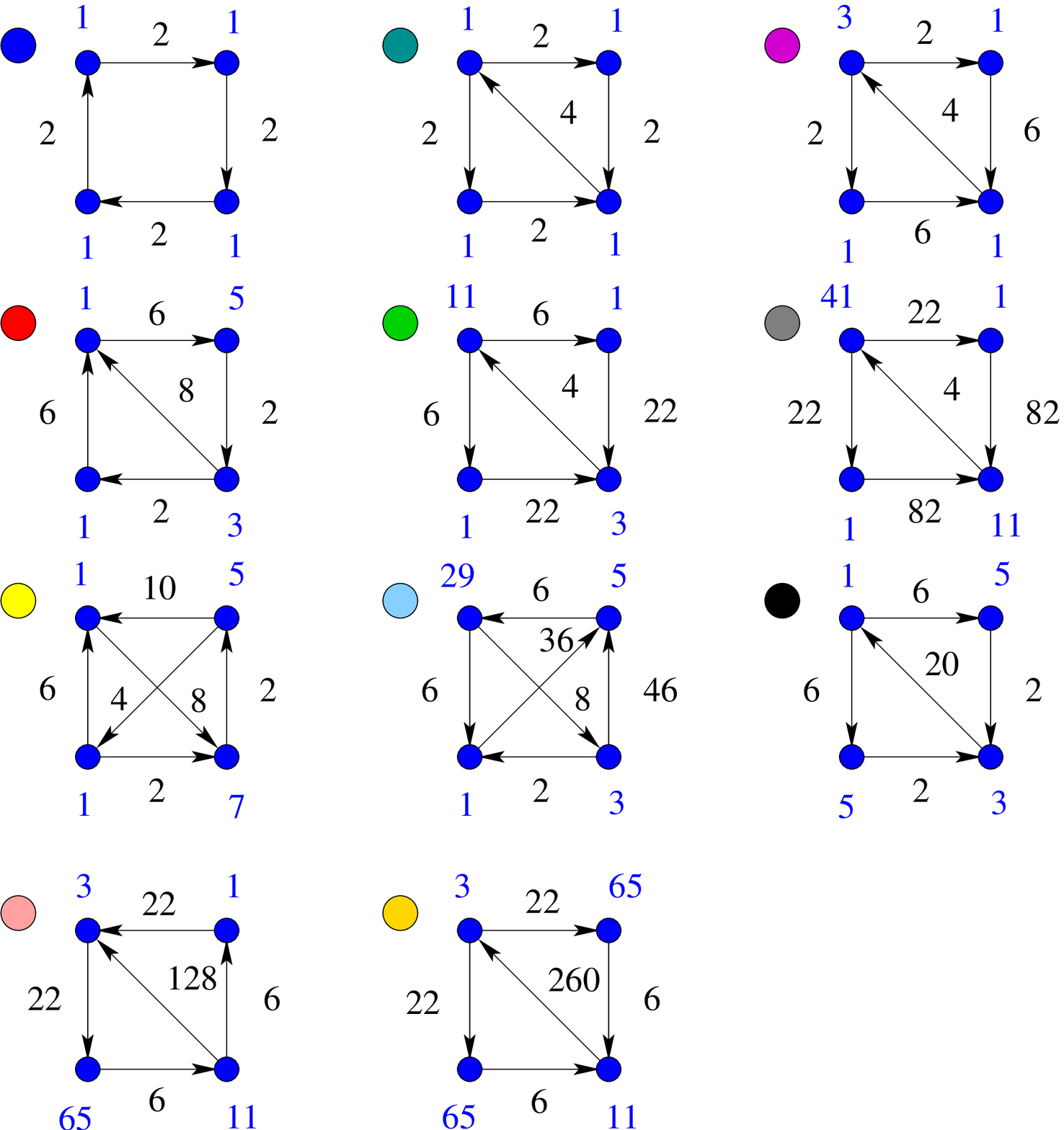} \ea
\label{F0}
\ea\eeq

There is a conifold-type cascade, given by the alternation between 2
models, which we will call A and B. Their quivers and the evolution of
the four inverse gauge couplings $x_i$
are given in \eref{F0-AB}:
\beq\ba{ccc}
\mbox{Model A} & \mbox{Model B}&\mbox{Running of Couplings} \\
  \ba{l} \epsfxsize = 2.5cm \epsfbox{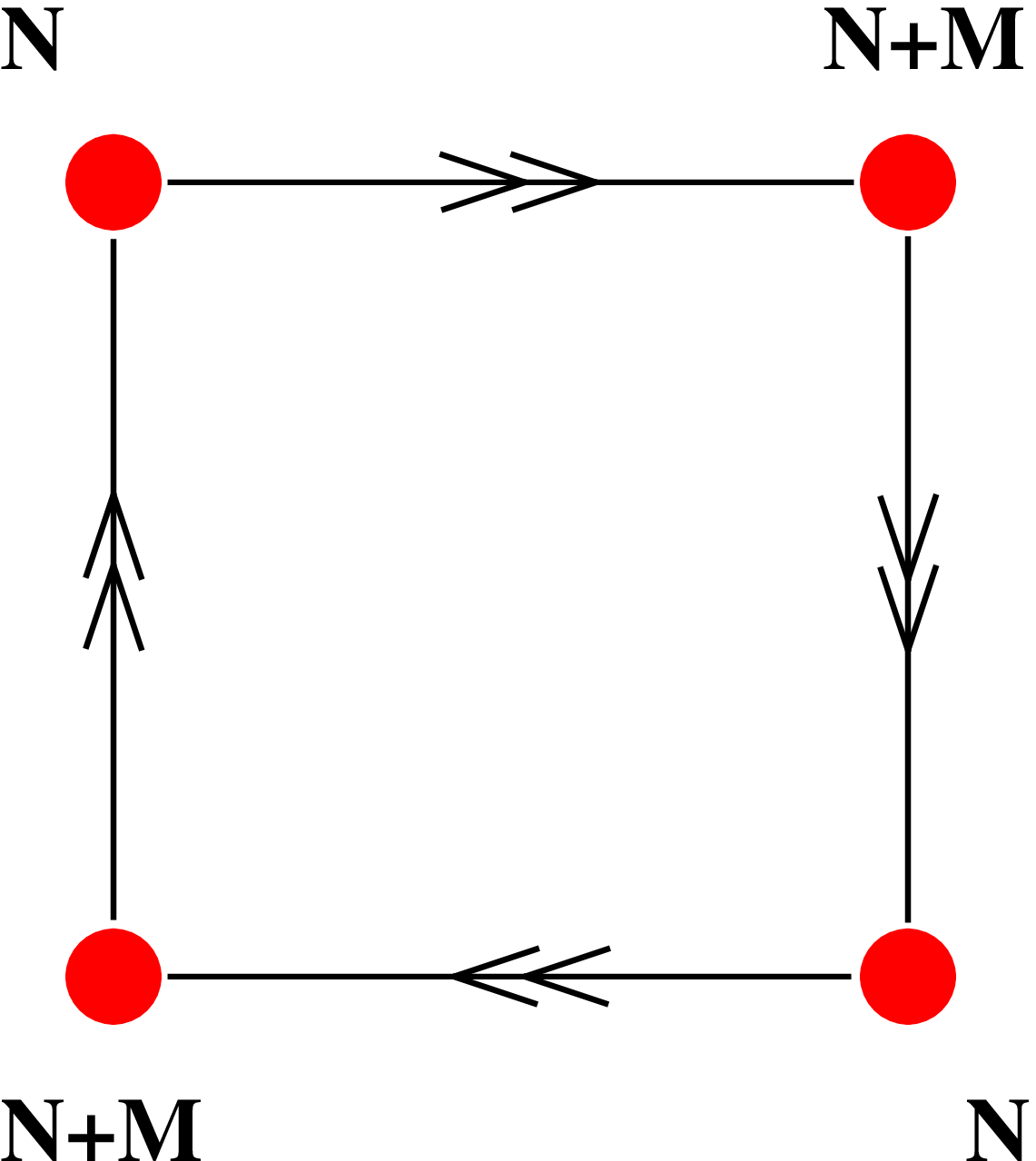} \ea
& \ba{l} \epsfxsize = 2.5cm \epsfbox{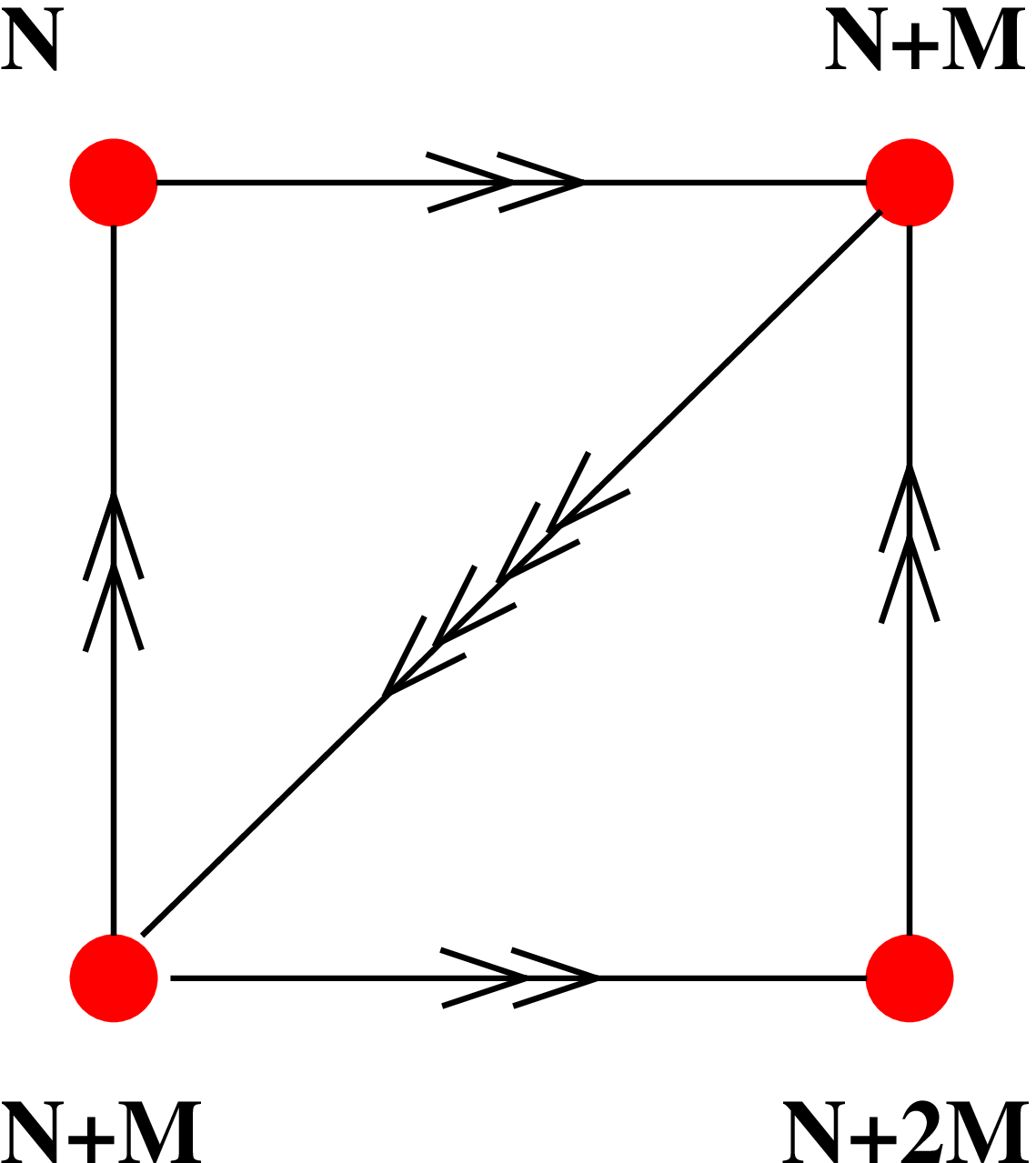} \ea
& \ba{r} \epsfxsize = 6cm \epsfbox{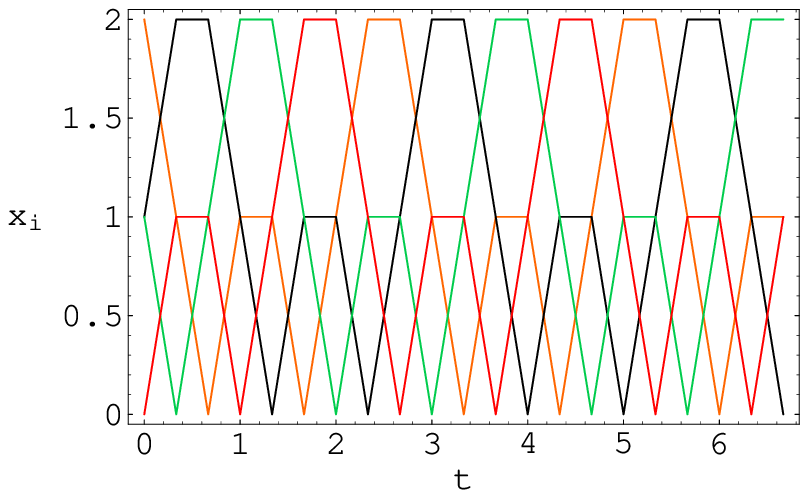}
\ea
\label{F0-AB}
\ea\eeq

This periodic RG cascade is represented as a closed cycle at the center of the 
duality tree in \eref{F0-AB}.

It is important to clarify at this point the philosophy that will be used along
the rest of the paper. We are interested in finding duality walls. As we described 
in the introduction, these phenomenon appears when reconstructing a duality cascade
that has a given theory at its IR bottom. Thus, we will proceed to derive one 
possible RG trajectory in the UV such that, when we consider the flow 
towards the IR using Seiberg duality in a cascading manner, we arrive at the 
desired model. The reader should keep in mind that this traditional perspective 
is considered although, for simplicity, we will number cascade steps starting from 
the IR theory and increasing towards the UV.

\subsection{The Duality Wall}

On the other hand, if we started model C (a Seiberg dual of models A and B), given in 
\eref{F0-C}, with associated beta-functions as shown,

\beq\ba{ccc}\label{F0-C}
  \ba{l} \epsfxsize = 2.5cm \epsfbox{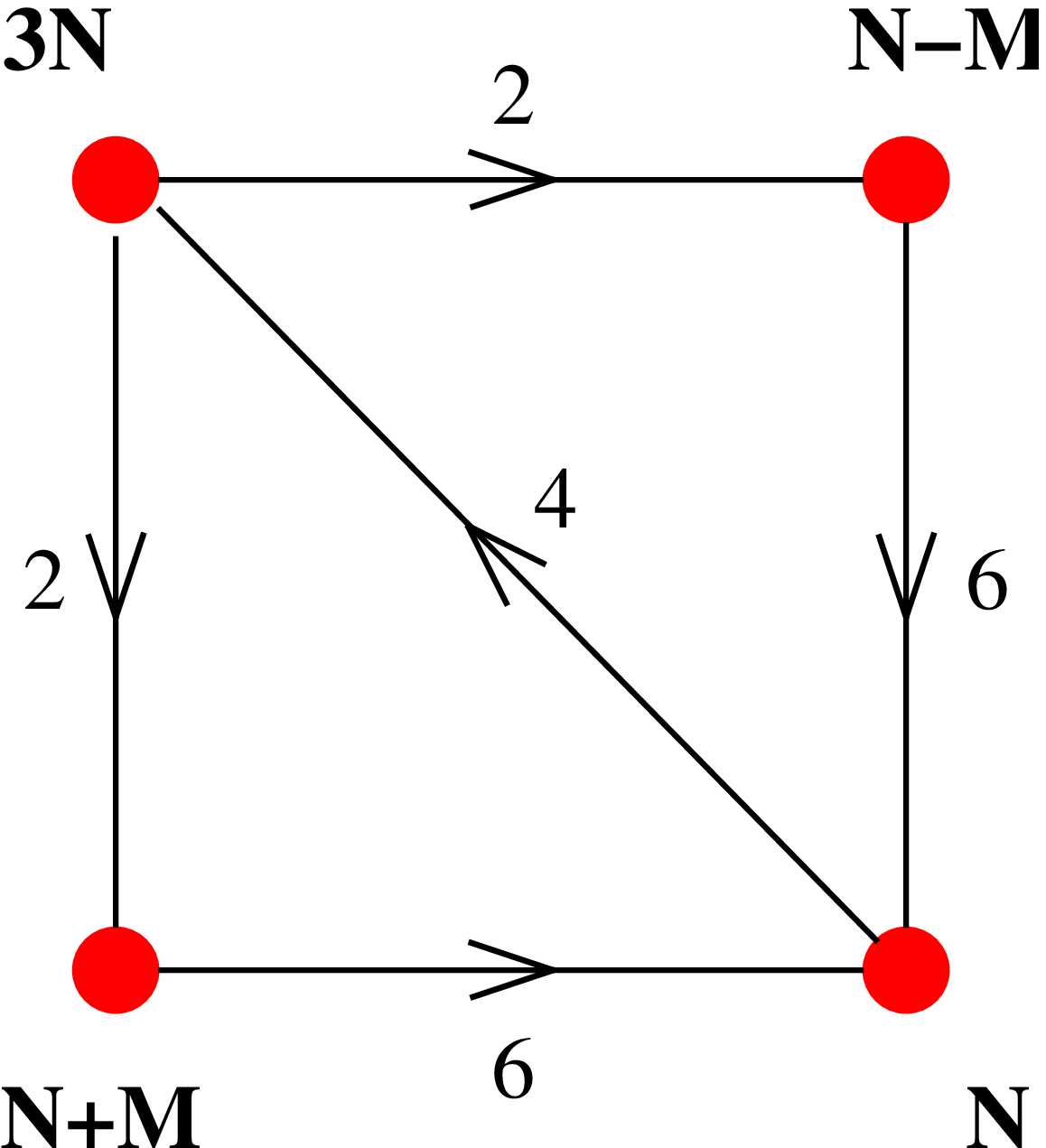} \ea
& \quad
& \ba{l} \beta_1=0 \\ \beta_2=-3M \\ \beta_3=0 \\ \beta_4=3M \ . \ea
\ea\eeq
a significantly different behavior emerges.

The inverse gauge couplings $x_i$, by definition \eref{betadef}, 
evolve with $\beta_i$ as
\beq
x_i(k) = x_i(k-1) + \beta_i(k-1) \Delta(k)
\eeq
for the $k$-th step in which the linear piece of the most negative
beta-function makes the corresponding $x$ step by $\Delta(k)$.
Let us, for concreteness, consider initial conditions
$(x_1,x_2,x_3,x_4)=(1,1,4/5,0)$. Then, the step $\Delta(k)$ decreases
at each dualization and the couplings $x_i$ evolve to a finite value:
\beq\ba{ccc}
\mbox{Evolution of Step Size} & &\mbox{Running of Couplings vs.~$t$} \\
\ba{l} \epsfxsize = 6cm \epsfbox{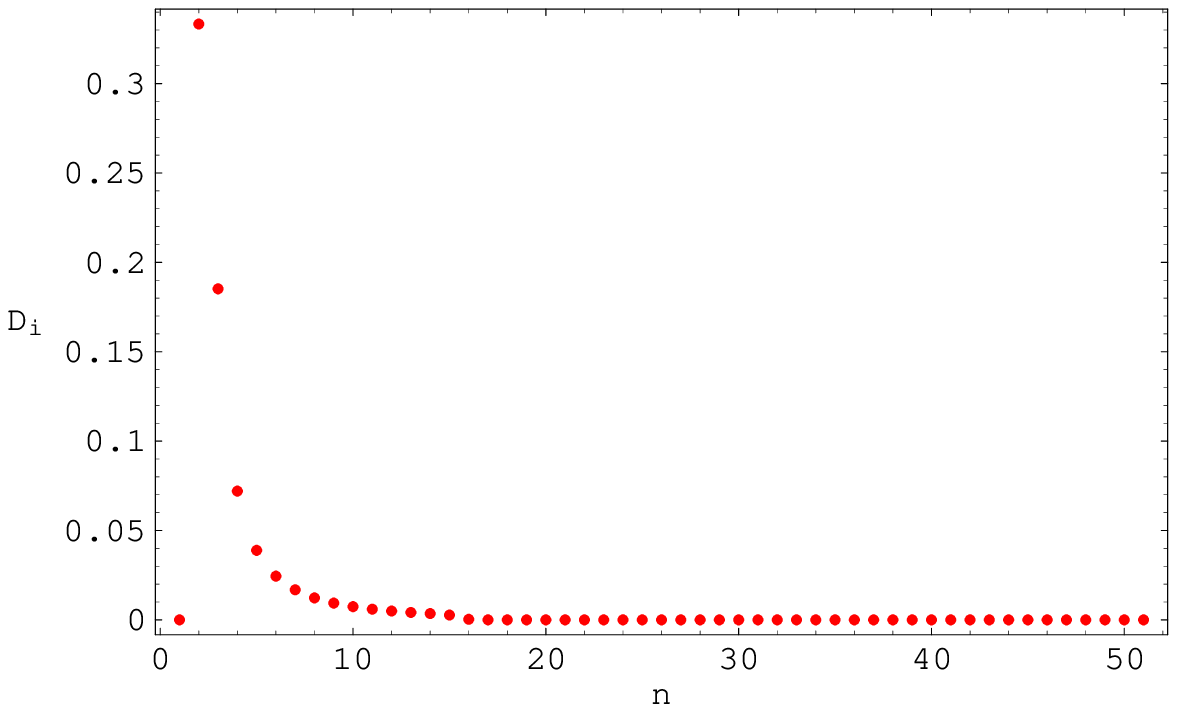} \ea
& \quad &
\ba{r} \epsfxsize = 6cm \epsfbox{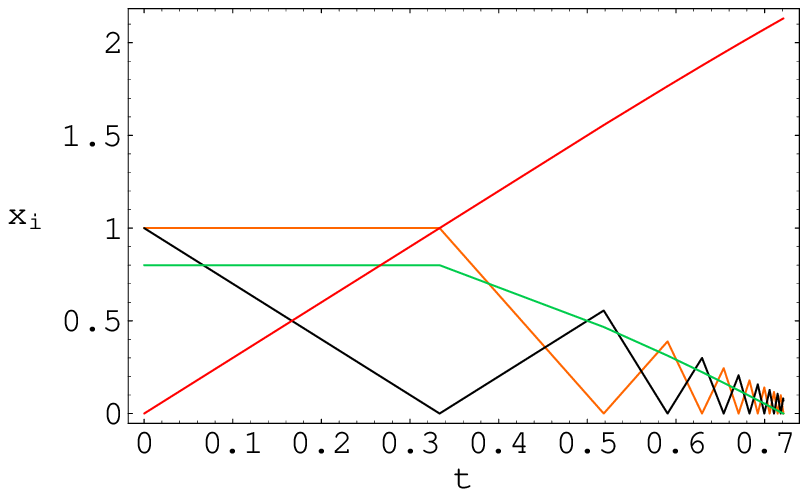} \ea
\ea\eeq
This is markedly different from the Klebanov-Strassler case.
Something very drastic occurs after the third node gets dualized, 
which produces an explosive growth of the number of chiral and vector 
multiplets in the quiver. After node 3 is dualized, the subsequent quivers 
have all their intersection numbers greater than 2 and become of hyperbolic type.
The cascade then is characterized by
a flow of the dualization scales towards an UV {\em accumulation
point} with a divergence in the number of degrees of
freedom. This phenomenon is called a {\it duality wall} \footnote{A similar
phenomenon has been observed in \cite{Ouyang}, in the context of supergravity.}.

\paragraph{A $\IZ_2$ symmetry as T-Duality}

Let us study how, starting from Model $C$, the RG flow 
continues further into the IR. This can be determined from our previous
results by simply changing the signs of the beta functions and replacing
the log-scale $t$ by $-t$. The later transformation, in the holographic dual,
resembles a T-duality-like action since the scale $\mu$ is associated with 
the radial coordinate. 
However, changing the sign of the beta function amounts to
changing $M$ to $-M$, which we observe to be
a $\IZ_2$ reflection of the quiver along the $(13)$ axis of the quiver
in \eref{F0-C}.
Therefore, the cascade to the IR from model C is simply the cascade that flows
into it from the UV after reflection of the quiver.

\paragraph{Sensitivity to initial conditions}

Let us briefly examine the sensitivity of the location of the duality wall 
to the initial inverse couplings. This problem was studied in \cite{Franco:2003ja},
and a detailed analytical study will appear in \cite{FHHW}.
Let our initial inverse gauge couplings be of the form $(1, x_2, x_3, 0)$, 
with $0< x_2,x_3 < 1$.
We study the running of the beta functions, and determine
the position of the duality wall, $t_{wall}$, for various initial values.
We plot in \eref{pos}, the position of the duality wall against
the initial values $x_2$ and $x_3$, both as a three-dimensional plot
in I and as a contour plot in II. We see that the position is a
step-wise function.
A similar behavior has been already observed in \cite{HW}.
\beq\label{pos}
\epsfxsize = 10cm\epsfbox{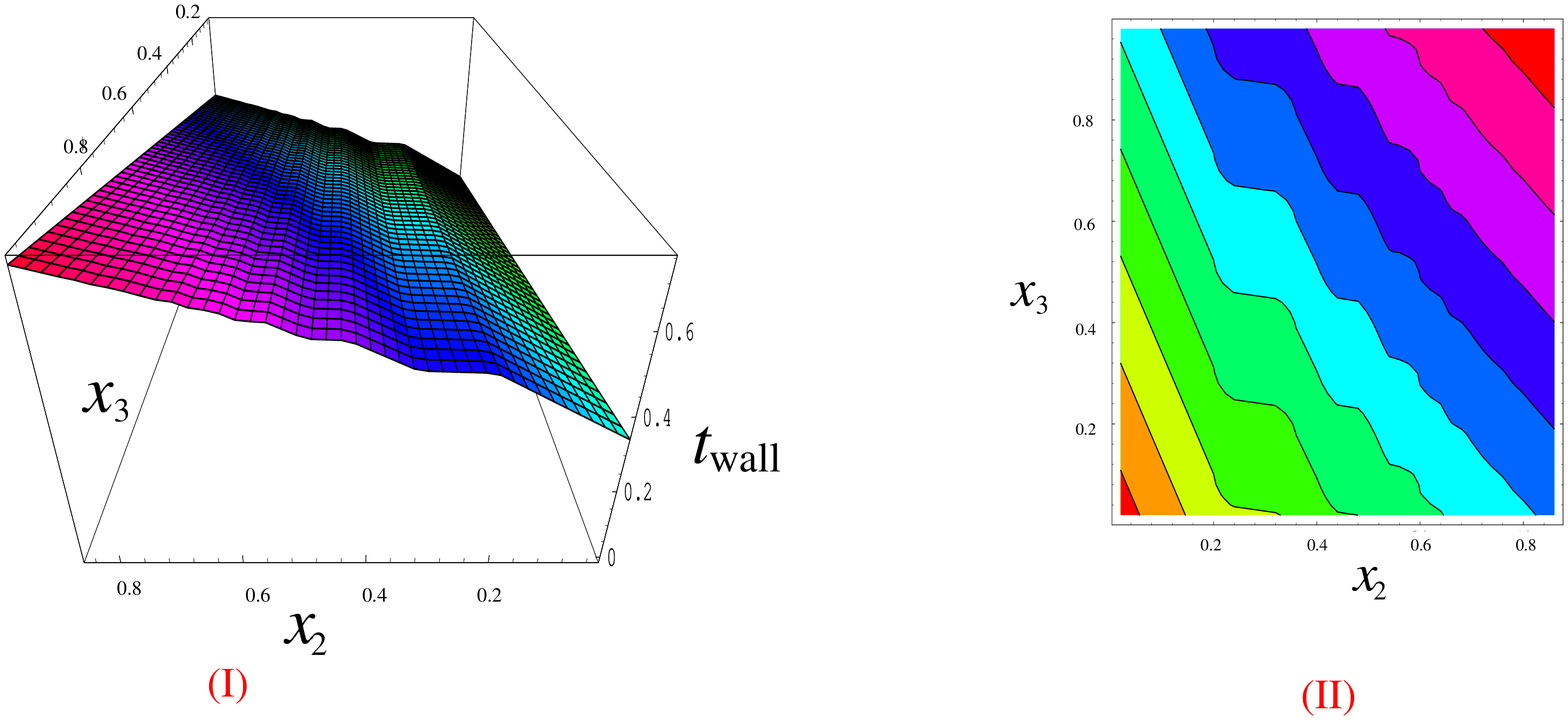}
\eeq

\section{Generating Cascades from Families of Diophatine Equations}
Having expounded upon the intricacies of duality cascades and flows, let
us return to some discussion on the generic structure of duality trees.
The connection between Seiberg Duality and Picard-Lefschetz (PL)
monodromy transformations 
(alternatively, mutations of exceptional collections) has been
explored in \cite{PL,conf1}. When  
studying different dual gauge theories that correspond to D-branes
over the same geometry,  
a quantity that remains invariant under PL transformations is the
trace of the total  
monodromy $K$ \cite{DeWolfe:1998pr}. For the
cases under study we have


\beq
\mbox{Tr} K=2
\eeq
This equation, when expressed in terms of the intersection numbers, gives
rise to a {\em Markov type Diophantine equation}. Such Diophatine equations 
have been derived for three-block exceptional collections over 
del Pezzos by algebraic geometers \cite{Karpov:1}. In
\cite{PL,chris}, the equations were shown to coincide. 
For a given singularity, there may be more than one such equations. 
For example, exceptional collections with
different number of blocks are 
classified by different Diophantine equations. 
In general, there may be more than one equation even for a given
number of blocks. Each of these equations 
encode an infinite set of all the gauge theories that
can be obtained for a given singularity using PL transformations, and
hence a subset of all Seiberg duals.
Remarkably, the set of 3-block equations for the del Pezzo surfaces 
is shown to be finite and all the equations can be organized into four
families \cite{Karpov:1} as follows
(the superscripts apply to those which
admit more than one Diophantine equation):
\beq
\begin{array}{rlcrl}
\mbox{{\bf Family I:}}   & dP_0, dP_6^I,dP_8^I & \ \ \ \ \ & 
\mbox{{\bf Family II:}} & F_0, dP_5, dP_7^I, dP_7^{II}, dP_8^{II} \\
\mbox{{\bf Family III:}} & dP_3, dP_6^{II}, dP_7^{III}, dP_8^{III} & &
\mbox{{\bf Family IV:}} & dP_4, dP_8^{IV} \ .
\end{array}
\label{Markov_eqs}
\eeq
Within each family, the equations are the same, up to a mere 
change of variables. In this way, we can start from a Diophantine
equation and its solutions for a given del Pezzo, and generate the
equation and solutions for other geometries. In other words, the
duality trees for different singularities may have subtrees that
coincide, and the map between them can be derived from the corresponding
equations.

Indeed, the physical content of this connection between Diophantine
equations makes it more than a quiver generating tool along a tree for
a given singularity. It is possible to
go beyond and use it to, starting from an RG flow containing solutions to
one of the members, generate entire cascades for other theories in the
same family.  Let us mention an explicit example to 
illustrate this statement. We have described in Section
\ref{section_F0_cascade} a KS type  
cascade for $F0$ with a flow involving a 3-block quiver (Model B). 
We can use \eref{Markov_eqs} to conclude that
there are analogous flows for $dP_5$, $dP_7$ and $dP_8$, the other
members of Family II. Furthermore 
the correspondence between theories in a given family indicates how to
choose fractional branes in order to obtain the related cascades.
We refer the reader to \cite{FHHW}, where these
cascades will be analyzed in detail.
It is natural to conjecture that this reasoning can be extended to
exceptional collections with a larger number of blocks
and their associated Diophantine equations and cascades.

\newpage

\section*{Acknowledgements}

It is a pleasure to thank Pavlos Kazakopoulos for collaboration in
the material presented. We would like to thank Francis Lam for many 
enlightening discussions and also C.~Herzog, U.~Gursoy,
C.~Nu\~{n}ez, M.~Schvellinger, and J.~Walcher for insightful
comments. S.F. appreciates the hospitality of the Institute for Advanced 
Study during the preparation of this work.
We sincerely acknowledge the gracious patronage of
the CTP and the LNS of MIT as well as
the department of Physics at UPenn. Further support is
granted from the U.S. Department of Energy under cooperative
agreements $\#$DE-FC02-94ER40818 and $\#$DE-FG02-95ER40893.
A.~H.~is also indebted to the Reed Fund Award
and a DOE OJI award, and Y.-H.~H., also to an NSF Focused Research
Grant DMS0139799 for ``The Geometry of Superstrings.''



\end{document}